\documentstyle[psfig,12pt,intj]{article} 

\begin{document}

\setlength{\baselineskip}{0.9cm} 
\title{VACUUM STATE OF LATTICE GAUGE THEORY \\
WITH FERMIONS IN 2+1 DIMENSIONS}
\author{Hideyuki Abe \\
{\it Institute of Physics, University of Tokyo,} \\
{\it Komaba, Tokyo 153, Japan}}
\date{}
\maketitle
\begin{abstract}
We investigate the vacuum state of the lattice gauge theory
with fermions in 2+1 dimensions.
The vacuum in the Hermite form for the fermion part is obtained;
the vacuum in the unitary form has been proposed by Luo and Chen.
It is shown that the Hermite vacuum has a lower energy
than the unitary one through the variational method.
\end{abstract}
\section{Introduction}

Quantum chromodynamics (QCD) is accepted
as the model of the strong interaction.
The lattice gauge theory is useful
for understanding the low energy behavior of QCD.
There are mainly two methods for formulating
the lattice gauge theory.
One method is the path integral formulation
in the Euclidean space--time.$^\cite{Wilson}$
In this formalism, the fermion part
of the QCD action is integrated into the fermion determinant.
Essentially, effects of the fermion appear
through the fermion determinant.
Its estimation is investigated in various ways.
Another method is the Hamiltonian formulation
in the Minkowski space.$^{\cite{KS},\cite{Banks}}$
In this formalism, the main effort has been made
for the pure gauge theory.
The vacuum wave function is constructed
in various ways.$^{\cite{Grnst},\cite{GZL}}$
It is expressed in terms of the loops of the link variables.
Moreover, the excited state orthogonal to the vacuum,
i.e.\ the glueball state,
is constructed.
One desires to understand the hadron dynamics analytically
rather than numerically.
However, there have been few attempts to construct
the vacuum with fermions.
Luo and Chen have investigated the vacuum
assuming the unitary form for the fermion field
with the variational method.$^\cite{LC}$

In this paper, we investigate the vacuum state
of the lattice gauge theory in 2+1 dimensions.
The lattice gauge theory in the (2+1)-dimensional space--time
is often dealt with since the (2+1)-dimensional theory is simple.
We expect that the results derived from the three-dimensional theory
can be generalized to the (3+1)-dimensional theory
for the most part.

In Sec.\ 2,
the vacuum state of the lattice gauge theory with the fermion field
is obtained in the strong coupling expansion.
The essential idea is that the state $ e^R|0\rangle $
is the vacuum, at least an eigenstate,
if the Hamiltonian is expressed as
$ H=e^{-R}H_{\rm eff}e^{-R}+E_0 $ in terms of the energy shift $ E_0 $
and certain operators $ R $ and $ H_{\rm eff} $,
which annihilates the strong coupling vacuum $ H_{\rm eff}|0\rangle=0 $.
In Sec.\ 3, we justify the strong coupling vacuum
through the $ t $ expansion method,
a useful tool for the Hamiltonian formulation.$^\cite{HW}$
In Sec.\ 4, we show that this vacuum has a lower energy expectation value
than that proposed by Luo and Chen.
A discussion is given in Sec.\ 5.
\section{Strong Coupling Vacuum}

We start with the latticized Hamiltonian in 2+1 dimensions
\begin{equation}
H=H_G+H_F.
\label{eq:H}
\end{equation}
The gauge part of the Hamiltonian, $ H_G=G+P $, is written as
\begin{eqnarray}
G & = & \frac{g^2}{2}\sum_{x} E_i^a(x)E_i^a(x), \\
P & = & -\frac{1}{2g^2}\sum_{x}\sum_{ij}
   {\rm tr}\left[U_{ij}(x)+U^\dagger_{ij}(x)\right],
\label{eq:HG}
\end{eqnarray}
where $ g $ is the dimensionless coupling constant.
In the Hamiltonian formulation, only spatial variables
are discretized.
The lattice spacing is assumed to be 1.
The link variable $ U_i(x) $ is defined
between the two adjacent sites $ x $ and $ x+\hat{i} $.
The plaquette is defined by
\begin{equation}
U_{ij}(x)=U_i(x)U_j(x+\hat{i})
    U^\dagger_i(x+\hat{j})U^\dagger_j(x).
\end{equation}
We use the four-spinor formulation
for the fermion field in three dimensions.$^\cite{ABKW}$
To avoid the doubling problem,
we adopt the staggered fermion,$^\cite{KS}$ $ H_F=T+mH_0 $:
\begin{eqnarray}
T & = & \frac{i}{2}\sum_{x}\eta_i(x)
   \left[\chi^\dagger(x)U_i(x)\chi(x+\hat{i})
   -\chi^\dagger(x+\hat{i})U^\dagger_i(x)\chi(x)\right], \\
H_0 & = & \sum_{x}(-1)^x\chi^\dagger(x)\chi(x),
\label{eq:HF}
\end{eqnarray}
where we have used the notations
$ \eta_i(x)=\delta_{i1}+\delta_{i2}(-1)^{x^1} $
and $ (-1)^x=(-1)^{x^1+x^2} $.
The (anti)commutators between the variables are defined by
\begin{equation}
[E_l^a,U_{l'}^{AB}]=-(T^aU_l)_{AB}\delta_{l,l'},
\label{eq:EUcom}
\end{equation}
\begin{equation}
\{\chi_A(x),\chi^\dagger_B(y)\}
  =\delta_{AB}\delta_{xy},
\end{equation}
where $ l $ denotes a link.
We use the $ \gamma $ matrices in the representation
$ \gamma^0=\sigma_3\otimes\sigma_3 $,
$ \gamma^0\gamma^i=-{\bf 1}\otimes\sigma_i $.
The relation between the fundamental fermion
and the Dirac spinor is given by
\begin{equation}
\psi_{2\alpha+\beta-2}(X) = \frac{1}{2\sqrt{2}}
 \sum_{\rho_1,\rho_2=0,1}
 \left(\sigma_1^{\rho_1}\sigma_2^{\rho_2}\right)_{\beta\alpha}
 \chi(2X+\rho),
\label{eq:DFrel}
\end{equation}
where the index of the Dirac spinor $ 2\alpha+\beta-2 $
runs from 1 to 4
because $ \sigma_1^{\rho_1}\sigma_2^{\rho_2} $
is a 2$ \times $2 matrix.
We proceed to construct the vacuum state.
We assume that $ g^2\gg m\gg 1/g^2 $.
The vacuum state of the pure gauge theory is developed
in various ways.
In one method,$^\cite{GZL}$
the vacuum state is given by $ |\tilde{0}\rangle=e^R|0\rangle $
if the Hamiltonian is expressed as
\begin{equation}
H = e^{-R}H_{\rm eff}e^{-R},
\label{eq:Heff}
\end{equation}
in the Hermitian operators $ R $ and $ H_{\rm eff} $,
which annihilates the strong coupling vacuum $ H_{\rm eff}|0\rangle=0 $.
To obtain the vacuum state with the fermion field,
we consider the strong coupling $ g\rightarrow\infty $
and heavy quark limit $ m\rightarrow\infty $.
The reason why we consider the heavy quark limit is
that the vacuum expectation value of the link variable,
contained by the fermion kinetic term, vanishes
in the strong coupling limit or in the random phase.
From the time evolution, $ \chi(x,t) $
is the annihilation operator of the particle
on the site $ x $ when $ (-1)^x=1 $,
and the production operator of the antiparticle
on the site $ x $ when $ (-1)^x=-1 $.
The vacuum state of the strong coupling
and heavy quark Hamiltonian is defined by
\begin{mathletters}
\begin{eqnarray}
E_l^a|0\rangle & = & 0, \\
\chi(x)|0\rangle & = & 0 \quad {\rm for} \quad(-1)^x=1, \\
\chi^\dagger(x)|0\rangle & = & 0 \quad {\rm for}\quad(-1)^x=-1.
\end{eqnarray}
\label{eq:scVacDef}
\end{mathletters}

Now, we express the Hamiltonian in the form (\ref{eq:Heff}).
We adopt the expression
\begin{eqnarray}
H & = & \frac{g^2}{2}\sum_{x}
   e^{-R}E_i^a(x)e^Re^RE_i^a(x)e^{-R} \nonumber \\
& & +h_1\sum_{(-1)^x=1}e^{-R}\chi^\dagger(x)e^Re^R\chi(x)e^{-R}
    \nonumber \\
& & +h_1\sum_{(-1)^x=-1}e^{-R}\chi(x)e^Re^R\chi^\dagger(x)e^{-R}
    \nonumber \\
& & +h_2\sum_{(-1)^x=1}e^{-R}\chi^\dagger_{A}(x)
        \chi_{B}(x+\hat{i})e^R
    e^R\chi^\dagger_{B}(x+\hat{i})
        \chi_{A}(x)e^{-R} \nonumber \\
& & +h_2\sum_{(-1)^x=-1}e^{-R}\chi^\dagger_{B}(x+\hat{i})
        \chi_{A}(x)e^R
    e^R\chi^\dagger_{A}(x)
        \chi_{B}(x+\hat{i})e^{-R}+E_0,
\label{eq:Hlast}
\end{eqnarray}
where the parameters $ h_1 $, $ h_2 $, and $ E_0 $
are to be determined.
For the pure gauge case,
only the first term is needed out of the Hamiltonian.
We have to add the four-fermion term to the expression (\ref{eq:Hlast})
because the term without link variables $ H_1 $
appears from the electric part:
\begin{equation}
H_1=\sum_{(-1)^x=1}\chi^\dagger_{A}(x)\chi_{B}(x+\hat{i})
    \chi^\dagger_{B}(x+\hat{i})\chi_{A}(x)
 +\sum_{(-1)^x=-1}\chi^\dagger_{B}(x+\hat{i})\chi_{A}(x)
    \chi^\dagger_{A}(x)\chi_{B}(x+\hat{i}).
\end{equation}
We expect that $ R $ is expressed as
\begin{eqnarray}
R & = & r_1R_1+r_{2a}R_{2a}+r_{2b}R_{2b}+r_{2c}R_{2c}
    +r_{2d}R_{2d}+R_g+{\rm O}(\frac{1}{g^6}).
\label{eq:Rexp}
\end{eqnarray}
The subscript in $ R_n $ stands for the operator
which contains $ n $ link variables.
We have already known the gauge part $ R_g $
investigated in various papers.
The lowest term of $ R_g $ is the plaquette
in the strong coupling expansion:
\begin{equation}
R_g = -\frac{1}{2g^2C(N_c)}P+{\rm O}(\frac{1}{g^8}),
\end{equation}
where $ C(N_c)=\frac{N_c^2-1}{2N_c} $ is a Casimir invariant.
We should note that we do not impose the plaquette ansatz
investigated in Ref.\ \cite{GZL} on the vacuum.
The other operators in Eq.\ (\ref{eq:Rexp}) are expressed as
$ R_1=T $ and
\begin{mathletters}
\begin{eqnarray}
R_{2a} & = & \frac{N_c+1}{2N_c}\sum_{x}
  \Bigl[ \left(\chi^\dagger(x)U_i(x)\chi(x+\hat{i})\right)^2
  +\left(\chi^\dagger(x+\hat{i})U^\dagger_i(x)\chi(x)\right)^2
  \Bigr], \\
R_{2b} & = & \frac{1}{N_c}\sum_{x}:\chi^\dagger(x)U_i(x)\chi(x+\hat{i})
  \cdot\chi^\dagger(x+\hat{i})U^\dagger_i(x)\chi(x):, \\
R_{2c} & = & \sum_{x}\Bigl[E_i^a(x),-\sum_j\eta_j(x-\hat{j})\eta_i(x)
     \Bigl(\chi^\dagger(x-\hat{j})U_j(x-\hat{j})
       T^aU_i(x)\chi(x+\hat{i}) \nonumber \\
& &  \qquad+\chi^\dagger(x+\hat{i})U^\dagger_i(x)
       T^a U^\dagger_j(x-\hat{j})\chi(x-\hat{j})
     \Bigr) \nonumber \\
& & +\sum_{j\neq i}\eta_j(x)\eta_i(x)
     \Bigl(\chi^\dagger(x+\hat{j})U^\dagger_j(x)
       T^aU_i(x)\chi(x+\hat{i}) \nonumber \\
& &  \qquad+\chi^\dagger(x+\hat{i})U^\dagger_i(x)
       T^aU_j(x)\chi(x+\hat{j})
     \Bigr) \nonumber \\
& & +\sum_j\eta_i(x)\eta_j(x+\hat{i})
     \Bigl(\chi^\dagger(x)T^aU_i(x)
       U_j(x+\hat{i})\chi(x+\hat{i}+\hat{j}) \nonumber \\
& &  \qquad+\chi^\dagger(x+\hat{i}+\hat{j})
       U^\dagger_j(x+\hat{i})U^\dagger_i(x)T^a\chi(x)
     \Bigr) \nonumber \\
& & -\sum_{j\neq i}\eta_i(x)\eta_j(x+\hat{i}-\hat{j})
     \Bigl(\chi^\dagger(x)T^aU_i(x)
       U^\dagger_j(x+\hat{i}-\hat{j})\chi(x+\hat{i}-\hat{j})
         \nonumber \\
& &  \qquad+\chi^\dagger(x+\hat{i}-\hat{j})
       U_j(x+\hat{i}-\hat{j})U^\dagger_i(x)T^a\chi(x)
     \Bigr) \Bigr], \\
R_{2d} & = & 2\sum_{x}\Bigl[E_i^a(x),-\chi^\dagger(x)T^a\chi(x)
    +\chi^\dagger(x+\hat{i})U^\dagger_i(x)
       T^aU_i(x)\chi(x+\hat{i})\Bigr],
\end{eqnarray}
\end{mathletters}
where the normal ordering is taken
in the meaning of Eqs.\ (\ref{eq:scVacDef}).
The graph of the operator $ R_1 $,
which implies the creation (or annihilation) of the quark pair,
is shown in Fig.\ 1(a).
The operators $ R_{2a} $, $ R_{2b} $, $ R_{2p} $
imply the creation (or annihilation) of the two quark pairs,
the creation after the annihilation of the quark pair,
and the creation (or annihilation) of the plaquette
respectively, shown in Figs.\ 1(b)--1(d).
\begin{figure}
\begin{center}
\setlength{\unitlength}{1in}
\begin{picture}(3.5,3)
\put(0,0){\makebox(3.5,3){\psfig{figure=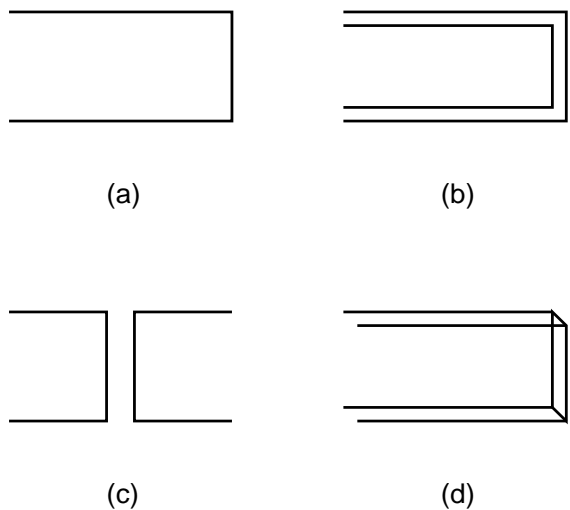,width=3.5in,height=3in}}}
\end{picture}
\end{center}
\caption{Graphs of the operators:
(a) the creation of the quark pair $ R_1 $,
(b) the creation of the two quark pairs $ R_{2a} $,
(c) the reproduction of the quark pair $ R_{2b} $,
(d) the creation of the plaquette $ R_{2p} $.}
\end{figure}
By requiring the equality
of Eqs.\ (\ref{eq:H}) and (\ref{eq:Hlast}) order by order,
we find the parameters to be
\begin{mathletters}
\begin{equation}
h_1 = m+\frac{4N_c}{N_c^2-1}\frac{1}{g^2}
   +{\rm O}(\frac{1}{g^4}),
\end{equation}
\begin{equation}
h_2 = -\frac{2N_c}{(N_c-1)(N_c^2-1)}\frac{1}{g^2}
   +{\rm O}(\frac{1}{g^4}),
\end{equation}
\begin{equation}
r_1 = \frac{2}{g^2C(N_c)}\left[1-\frac{4m}{g^2C(N_c)}\right]
 +{\rm O}(\frac{1}{g^6}),
\end{equation}
\begin{equation}
r_{2a} = -\frac{N_c}{N_c^2-N_c-2}\left[\frac{2}{g^2C(N_c)}\right]^2
 +{\rm O}(\frac{1}{g^6}),
\end{equation}
\begin{equation}
r_{2b} = -\frac{1}{N_c}\left[\frac{2}{g^2C(N_c)}\right]^2
 +{\rm O}(\frac{1}{g^6}),
\end{equation}
\begin{equation}
r_{2c} = \frac{1}{2C(N_c)}\left[\frac{2}{g^2C(N_c)}\right]^2
 +{\rm O}(\frac{1}{g^6}),
\end{equation}
\begin{equation}
r_{2d} = \frac{1}{2N_c}\left[\frac{2}{g^2C(N_c)}\right]^2
 +{\rm O}(\frac{1}{g^6}),
\end{equation}
\begin{equation}
E_0 = -\frac{4N_c^2}{N_c^2-1}\frac{1}{2g^2}L^2
    +{\rm O}(\frac{1}{g^4}),
\end{equation}
\label{eq:Rparam}
\end{mathletters}
for sufficiently large $ N_c $.
The number of the lattice sites is assumed
to be $ L^2=\sum_{x} 1 $.
The deviation of the parameter $ h_1 $
from the quark mass $ m $
shows the mass generation in the effective Hamiltonian.
The Hamiltonian (\ref{eq:Hlast}) reproduces
the normal-ordered form of the expression (\ref{eq:H})
up to order $ 1/g^2 $.
Consequently, the vacuum state
is $ |\tilde{0}\rangle=e^R|0\rangle $.
It is an eigenstate of the Hamiltonian
\begin{equation}
H|\tilde{0}\rangle=E_0|\tilde{0}\rangle.
\label{eq:eignstt}
\end{equation}

We find that the pseudoscalar meson
$ \psi^\dagger\gamma^0\gamma^5\psi|\tilde{0}\rangle $
and the vector meson
$ \psi^\dagger\gamma^0\gamma^1\psi|\tilde{0}\rangle $
with zero momentum take the same energy expectation value,
\begin{equation}
E=\frac{g^2}{2}C(N_c)+2m+E_0
  +\frac{1}{g^2}\frac{2}{C(N_c)},
\end{equation}
where we have used the additional $ \gamma $ matrix defined by
$ \gamma^5=-\sigma_2\otimes{\bf 1} $.
\section{Comparison with $ t $ Expansion}

In the previous section, we have obtained the vacuum
in the strong coupling expansion.
It is instructive to compare our vacuum
with that obtained by the $ t $ expansion.
The $ t $ expansion vacuum is given
as a trial state with the damping factor $ e^{-Ht} $
in the limit $ t\rightarrow\infty $.
The real vacuum is the lowest energy state
that survives in this limit:
\begin{equation}
\lim_{t\rightarrow\infty}\frac{e^{-\bar{H}t}|0\rangle}{\langle 0|e^{-2\bar{H}t}|0\rangle^{1/2}}
    =\frac{|\tilde{0}\rangle}{\langle\tilde{0}|\tilde{0}\rangle^{1/2}},
\end{equation}
where we have used the Hamiltonian divided by $ g^2 $,
$ \bar{H}=H/g^2 $, for the convenience of the order counting.
We also use the notations $ \bar{G} $, $ \bar{P} $,
$ \bar{T} $, and $ \bar{H_0} $
for the electric part, the plaquette term,
the fermion kinetic term, and the fermion mass term
divided by $ g^2 $ respectively.
Since our vacuum is given in the strong coupling expansion,
we use a spoiled version of the $ t $ expansion,
{\it i.\ e}.\ we do not apply the continuation to extract the physical value,
Pad$ \acute{{\rm e}} $ approximation, etc.
The terms multiplied by $ (\bar{G}+m\bar{H_0})^n $ for all $ n $
should be considered together
because its eigenvalue is usually of order $ g^0 $.
Out of the $ t $ expansion vacuum,
the terms proportional to $ |0\rangle $ are
\begin{eqnarray}
|0'\rangle & = &
 |0\rangle+\left(\frac{1}{2g^2}\right)^2\sum_l\sum_{n=2}^\infty
 \frac{(-t)^n}{n!}N_c\left(\bar{G}+m\bar{H_0}\right)^{n-2}
    |0\rangle.
\end{eqnarray}
These terms are almost from the expansion
of the exponential function
\begin{eqnarray}
|0'\rangle
& = & \left[1+\frac{1}{4}N_c\frac{1}{g^4}C^{-2}
  \left(e^{-Ct}-1+Ct\right)2L^2\right]|0\rangle,
\end{eqnarray}
where $ C=C(N_c)/2+2m/g^2 $
is the eigenvalue of $ \bar{G}+m\bar{H_0} $
for the quark pair state $ \bar{T}|0\rangle $.
The quark pair states, shown in Fig.\ 2, are assembled to
\begin{eqnarray}
|l\rangle
& = & \frac{1}{C}\left(e^{-Ct}-1\right)\bar{T}|0\rangle.
\end{eqnarray}
\begin{figure}
\begin{center}
\setlength{\unitlength}{1in}
\begin{picture}(2.5,1.8)
\put(0,0){\makebox(2.5,1.8){\psfig{figure=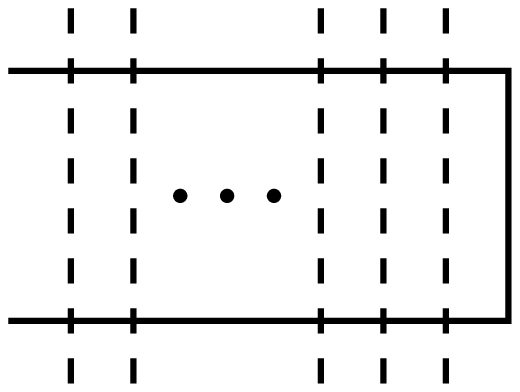,width=2.5in,height=1.8in}}}
\end{picture}
\end{center}
\caption{Graph of the quark pair state in $ t $ expansion.
Dashed lines denote the operator $ \bar{G}+m\bar{H_0} $.}
\end{figure}
The double quark pair states are summed to
\begin{eqnarray}
|ll'\rangle
& = & -\frac{1}{4g^4}\frac{1}{2C^2}
  \left(1+e^{-2Ct}-2e^{-Ct}\right)
  \sum_{l\neq l'}\eta_l\eta_{l'}\chi^\dagger U_l\chi
     \cdot\chi^\dagger U_{l'}\chi|0\rangle,
\end{eqnarray}
for $ l \neq l' $ and
\begin{eqnarray}
|ll\rangle
& = & -\frac{1}{4g^4}\frac{1}{1-r}
  \left(1-r+re^{-Ct}-e^{-Crt}\right)\frac{1}{rC^2}
  \sum_l\left(\chi^\dagger U_l\chi\right)^2|0\rangle,
\end{eqnarray}
for the same $ l $
where $ Cr=(N_c^2-N_c-2)/2N_c+4m/g^2 $
is the eigenvalue of $ \bar{G}+m\bar{H_0} $
for the state $ (\chi^\dagger U_l\chi)^2|0\rangle $.
The plaquette state is
\begin{eqnarray}
|P\rangle
& = & \left\{\exp\left[-2C(N_c)t\right]-1\right\}
   \frac{1}{2C(N_c)}\bar{P}|0\rangle.
\end{eqnarray}
The $ t $ vacuum state in the limit $ t\rightarrow\infty $
is consistent with the vacuum $ e^R|0\rangle $,
\begin{eqnarray}
\lefteqn{ \lim_{t\rightarrow\infty}
  \frac{\left[|0'\rangle+|l\rangle+|ll'\rangle
  +|ll\rangle+|P\rangle\right]}{\langle 0|e^{-2\bar{H}t}|0\rangle^{1/2}} } \nonumber \\
& = & \frac{\left[1+r_1R_1+r_{2a}R_{2a}+R_{2g}
   +{\textstyle\frac{1}{2}}r_1^2R_1^2\right]|0\rangle}{\langle\tilde{0}|\tilde{0}\rangle^{1/2}},
\end{eqnarray}
up to order $ 1/g^4 $.
\section{Variational Vacuum}

So far, we have investigated the vacuum in the strong coupling expansion.
Luo and Chen have proposed the vacuum in the unitary form for the fermion part.
The unitary vacuum contradicts the one we have obtained,
which we call the Hermite vacuum.
In this section,
we compare the two vacua through the variational method.
We show that a lower energy minimum is realized for the hermite vacuum
than for the unitary one.
To compare the two vacua on the equal condition,
we adopt the same form for the gauge part in both cases.
We use the one-plaquette formulation,$^\cite{GZL}$
in which the gauge part $ R_g $ is truncated up to the plaquette term,
because we can calculate the vacuum expectation value without variation.
For the fermion part,
we use the one-link approximation $ R_f=\theta_f T $ as a trial state
with a variational parameter $ \theta_f $
and the kinetic term of the fermion $ T $
defined in Eq.\ (\ref{eq:HF}).
We calculate the energy expectation value
for the Hermite vacuum $ \exp(R_f)\exp(R_g)|0\rangle $
and the unitary one, $ \exp(iR_f)\exp(R_g)|0\rangle $.
The Hamiltonian can effectively be written as
$ \langle H \rangle=\langle H_{\rm eff} \rangle $ under the trial state,
where
\begin{eqnarray}
H_{\rm eff} & = & \left[1+\frac{g^2}{2}C(N_c)\theta_f\right]T
+\left[m+\frac{g^2}{2}C(N_c)\theta_f^2\right]H_0
+\frac{g^2}{8}\theta_f^2(R_{2a}+R_{2b}-H_1) \nonumber \\
& & -\frac{g^2}{4}N_cC(N_c)L^2\theta_f^2, \\
H_{\rm eff} & = & \left[1+i\frac{g^2}{2}C(N_c)\theta_f\right]T
+\left[m-\frac{g^2}{2}C(N_c)\theta_f^2\right]H_0
-\frac{g^2}{8}\theta_f^2(R_{2a}+R_{2b}-H_1) \nonumber \\
& & +\frac{g^2}{4}N_cC(N_c)L^2\theta_f^2,
\end{eqnarray}
for the Hermite and unitary vacuum respectively.
Once the effective Hamiltonian is expressed in the link variable,
we can estimate the energy expectation value
using the one-plaquette formulation.
The link graph with crossings can be factorized
into bubbles and a diagram without crossings.
Now, the energy density $ {\cal E}=\langle H \rangle/(N_cL^2) $
is expressed as a function of $ \theta_f $,
\begin{eqnarray}
{\cal E}(\theta_f) & = & -\frac{g^2}{4}C(N_c)\theta_f^2
+\left[m+\frac{g^2}{2}C(N_c)\theta_f^2\right](A\theta_f+B) \nonumber \\
& & +\left[1+\frac{g^2}{2}C(N_c)\theta_f
  -\frac{g^2}{8}C(N_c)\theta_f^3\right]A, \\
{\cal E}(\theta_f) & = & \frac{g^2}{4}C(N_c)\theta_f^2
+\left[m-\frac{g^2}{2}C(N_c)\theta_f^2\right]B,
\end{eqnarray}
for the Hermite and the unitary vacuum respectively,
where $ A $ and $ B $ are polynomials of $ \theta_f $:
\begin{mathletters}
\begin{eqnarray}
A & = & \sum_{m,n}(-1)^{m+n-1}Y^{mn}
  \frac{(m+n)!(m+n)!}{(2m+2n-1)!}2\theta_f^{2m+2n-1}, \\
B & = & -4\sum_{m,n}(-1)^{m+n-1}Y^{mn}
  (m+n)\left(\frac{\theta_f}{2}\right)^{2m+2n}.
\end{eqnarray}
\label{eq:poly}
\end{mathletters}
The summation is taken over $ m,n $, the extension of the loop graph
in the direction 1,2.
We show the result for the SU(3) case below.
The Hermite vacuum has a lower energy minimum than the unitary one,
as in Fig.\ 3.
\begin{figure}
\begin{center}
\setlength{\unitlength}{1in}
\begin{picture}(4,2.6)
\put(0,0){\makebox(4,2.6){\psfig{figure=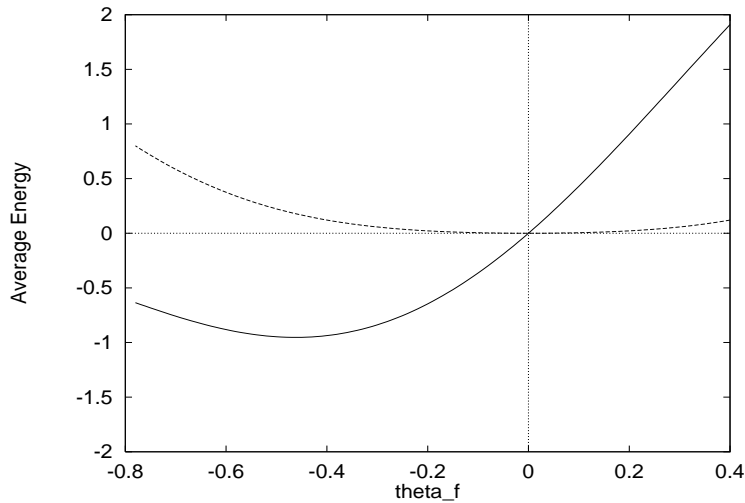,%
width=4in,height=2.6in,angle=-90}}}
\end{picture}
\end{center}
\caption{The average energy $ {\cal E}=\langle H\rangle/(N_cL^2) $
as a function of the variational parameter $ \theta_f $ at $ 1/g^2=0.7 $.
The solid curve for the Hermite vacuum
and the dashed curve for the unitary one.}
\end{figure}
The variational parameter $ \theta_f $ of the Hermite vacuum
takes a nontrivial value,
and that of the unitary vacuum zero in the one-plaquette approximation.
From the relation (\ref{eq:DFrel}),
the chiral condensate is expressed by the polynomials (\ref{eq:poly}):
\begin{eqnarray}
\langle\bar{\psi}\psi\rangle & = &
  \frac{N_c}{4}\left(-\frac{1}{2}+A\theta_f+B\right), \\
\langle\bar{\psi}\psi\rangle & = &
  \frac{N_c}{4}\left(-\frac{1}{2}+B\right),
\end{eqnarray}
for the Hermite and the unitary vacuum respectively.
The chiral condensate shows the scaling behavior
in the vicinity of $ 1/g^2=0.66$, as in Fig.\ 4.
\begin{figure}
\begin{center}
\setlength{\unitlength}{1in}
\begin{picture}(4,2.6)
\put(0,0){\makebox(4,2.6){\psfig{figure=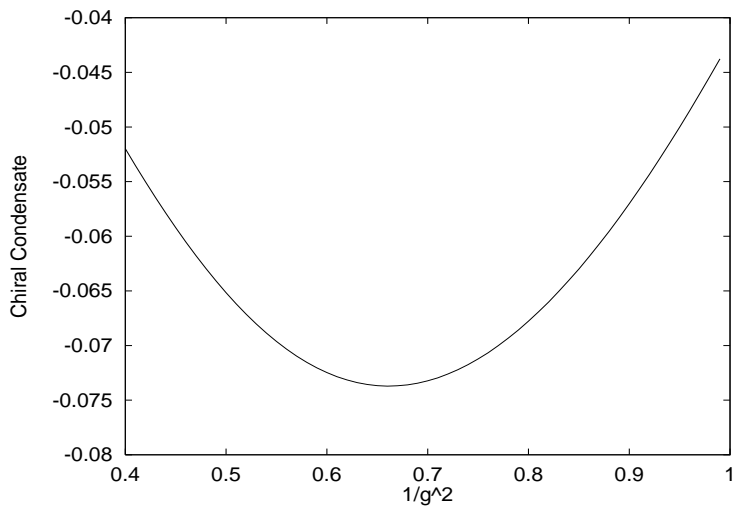,%
width=4in,height=2.6in,angle=-90}}}
\end{picture}
\end{center}
\caption{The scaled chiral condensate
$ \langle\bar{\psi}\psi\rangle/(g^4N_c) $
as a function of $ 1/g^2 $. The scaling behavior is observed
in the vicinity of $ 1/g^2=0.66 $.}
\end{figure}
The scaled chiral condensate $ \langle\bar{\psi}\psi\rangle/g^4 $
is estimated at $ -0.074\times 3 $.
To reach this value,
a higher order calculation is required for the unitary vacuum,
as in Ref.\ \cite{LC}.

It is important to show the mass of the pseudoscalar meson
as a Goldstone boson caused by the chiral symmetry breaking.
The pseudoscalar meson state is expressed by
$ {\rm exp}(R_f+R_g)a^\dagger|0\rangle $
in the same $ R_f $ but with the different parameter $ \theta_f $,
where the production operator in the strong coupling limit is
\begin{equation}
a^\dagger =\sum_x\Bigl[\chi^\dagger(2x)U_2(2x)\chi(2x+2)
+\chi^\dagger(2x+1+2)U_2^\dagger(2x+1)\chi(2x+1)\Bigr].
\end{equation}
The mass of the pseudo-scalar meson, orthogonalized to the vacuum,
is obtained as the energy difference from the vacuum energy
\begin{eqnarray}
M_{\rm PS} & = & \frac{1}{N}\sum_{m'n}(-1)^{n-1}Y^{2m'n}
 \left[\frac{n^2+1}{2}\right]\theta_f^{4m'+2n-3}
  \Biggl\{\left[m+\frac{g^2}{2}C(N_c)\theta_f^2\right]\theta_fC_{2m'n}^{(1)}
  \nonumber \\
& & +\left[1+\frac{g^2}{4}C(N_c)\theta_f^2\right]C_{2m'n}^{(2)}
-\frac{g^2}{4}C(N_c)\theta_f^3C_{2m'n}^{(3)}\Biggr\}+\frac{g^2}{2}C(N_c), \\
N &=& \sum_{m'n}(-1)^{n-1}Y^{2m'n}\left[\frac{n^2+1}{2}\right]
  \theta^{4m'+2n-2}D_{2m'n},
\end{eqnarray}
where the coefficients are defined by
\begin{mathletters}
\begin{eqnarray}
C_{mn}^{(1)} & = & 2(m+n)D_{mn}
  -(m+n-1)\left(\frac{1}{2}\right)^{2m+2n-1}, \\
C_{mn}^{(2)} & = & 2D_{mn}-\left(\frac{1}{2}\right)^{2m+2n-3}, \\
C_{mn}^{(3)} & = & (m+n-1)D_{mn}, \\
D_{mn} & = & \frac{(m+n-1)!(m+n-1)!}{(2m+2n-2)!}.
\end{eqnarray}
\end{mathletters}
The scaled mass is 0.61 at $ 1/g^2=0.66 $, as in Fig.\ 5.
The small value remains
owing to the explicit chiral symmetry breaking
to solve the doubling problem.
\begin{figure}
\begin{center}
\setlength{\unitlength}{1in}
\begin{picture}(4,2.6)
\put(0,0){\makebox(4,2.6){\psfig{figure=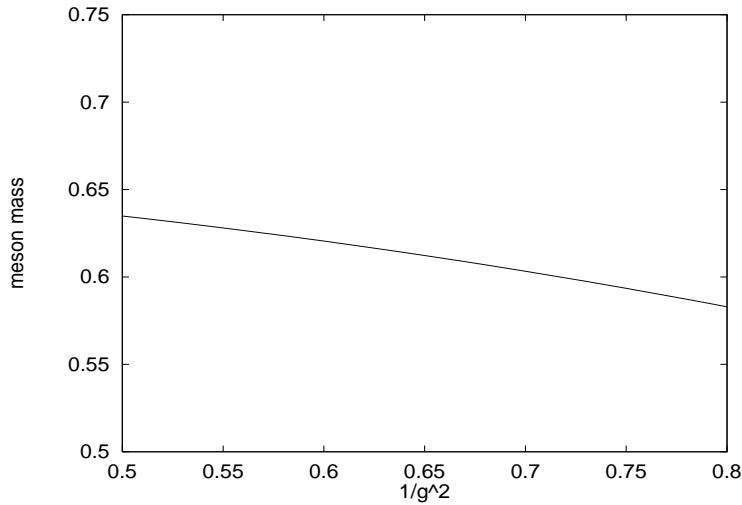,%
width=4in,height=2.6in,angle=-90}}}
\end{picture}
\end{center}
\caption{The scaled mass of the pseudoscalar meson
$ M_{\rm PS}/g^2 $ as a function of $ 1/g^2 $.}
\end{figure}
\section{Discussion}

We discuss our results in this section.
The vacuum state of the lattice gauge theory
with the fermion field is investigated.
The vacuum in the unitary form for the fermion part
proposed in Ref.\ \cite{LC} is different from the Hermite vacuum
obtained in the present paper.
There is no reason to adopt the vacuum in the unitary form.
A principle is required for the determination of the vacuum form.
We have proposed the vacuum form
inspired from the strong coupling expansion.
It is shown that the Hermite vacuum has a lower energy than the unitary one
through the variational method.

The chiral condensate of the Hermite vacuum in the one-link approximation
is consistent with that of the unitary one in the two-link approximation
in Ref.\ \cite{LC}.
This implies that the Hermite vacuum
has a better convergence than the unitary one,
in which only a trivial value is obtained in the one-link approximation.

The mass of the pseudoscalar meson is estimated.
Although its scaled value does not vanish completely,
it is consistent because the chiral symmetry is broken explicitly
to solve the doubling problem.

We should note that our formulation of the vacuum structure breaks down
in the strong coupling expansion for the SU(2) case.
The factor $ 1/m $ appears
in the coefficient of the term $ R_{2a} $.
Such a factor in the vacuum
suggests that another formalism is required in the SU(2) case.

\newpage
\pagestyle{empty}
\begin{figurecaptions}
\item Graphs of the operators:
(a) the creation of the quark pair $ R_1 $,
(b) the creation of the two quark pairs $ R_{2a} $,
(c) the reproduction of the quark pair $ R_{2b} $,
(d) the creation of the plaquette $ R_{2p} $.

\item Graph of the quark pair state in $ t $ expansion.
Dashed lines denote the operator $ \bar{G}+m\bar{H_0} $.

\item The average energy $ {\cal E}=\langle H\rangle/(N_cL^2) $
as a function of the variational parameter $ \theta_f $ at $ 1/g^2=0.7 $;
the solid curve for the Hermite vacuum
and the dashed curve for the unitary one.

\item The scaled chiral condensate $ \langle\bar{\psi}\psi\rangle/(g^4N_c) $
as a function of $ 1/g^2 $. The scaling behavior is observed
in the vicinity of $ 1/g^2=0.66 $.

\item The scaled mass of the pseudoscalar meson
$ M_{\rm PS}/g^2 $ as a function of $ 1/g^2 $.
\end{figurecaptions}

\end{document}